\documentclass[letterpaper, 10 pt, conference]{ieeeconf}  

\IEEEoverridecommandlockouts                              

\usepackage{amsmath,amssymb}
\usepackage{subfigure}
\usepackage{xcolor}

\newcommand{\cut}[1]{}
\newcommand{\E}{\mathbb{E}}

\newcommand{\R}{\mathbb{R}}

\newcommand{\upstairs}[1]{\textsuperscript{#1}}
\newcommand{\affilone}{\dag}
\newcommand{\affiltwo}{\ddag}

\usepackage{graphicx}
\graphicspath{ {./figures/} }

\title{\LARGE \bf
Analysis of the Impact of Mask-wearing in Viral Spread: Implications for COVID-19
}

\author{Yurun Tian\upstairs{1,\affilone},  Anirudh Sridhar\upstairs{2,\affiltwo}, Osman Ya\u{g}an\upstairs{3,\affilone}, H. Vincent Poor\upstairs{4,\affiltwo} \\
{ \upstairs{\affilone} Department of Electrical and Computer Engineering, Carnegie Mellon University} \\
{ \upstairs{\affiltwo} Department of Electrical Engineering, Princeton University}
\thanks{\upstairs{1}{\small yurunt@andrew.cmu.edu}, \upstairs{2}{\small anirudhs@princeton.edu}} \thanks{\upstairs{3}{\small oyagan@andrew.cmu.edu}, \upstairs{4}{\small poor@princeton.edu}}%
}

\begin{document}

\maketitle
\thispagestyle{empty}
\pagestyle{empty}

\begin{abstract}

Masks are used as part of a comprehensive strategy of measures to limit transmission and save lives during the COVID-19 pandemic. Research about the impact of mask-wearing in the COVID-19 pandemic has raised formidable interest across multiple disciplines. In this paper, we investigate the impact of mask-wearing in spreading processes over complex networks. This is done by studying a heterogeneous bond percolation process over a multi-type network model, where nodes can be one of two types (mask-wearing, and not-mask-wearing). We provide analytical results that accurately predict the expected epidemic size and probability of emergence as functions of the characteristics of the spreading process (e.g., transmission probabilities, inward and outward efficiency of the masks, etc.), the proportion of mask-wearers in the population, and the structure of the underlying contact network. In addition to the theoretical analysis, we also conduct extensive simulations on random networks. We also comment on the analogy between the mask-model studied here and the multiple-strain viral spreading model with mutations studied recently by Eletreby et al.

\end{abstract}

\section{INTRODUCTION}
The rapid spread of COVID-19 has devastated the world since its inception in December 2019, leading to global economic crises and claiming hundreds of thousands of lives. As schools and businesses reopen, it is of paramount importance to asses how various safety measures may limit the spread of COVID-19. One such measure is mask-wearing, which is known to reduce the transmissibility of viruses that spread through respiratory droplets. Much of the existing work surrounding the effectiveness of mask-wearing have studied how it limits transmission between individuals \cite{mask_wearing1,mask_wearing2}. However, several questions remain regarding the health of the general public. How many people must wear masks to significantly curb the spread of COVID-19? More generally, how does mask-wearing change the spreading dynamics of an epidemic? 

In this paper, we provide quantitative answers to the questions above. To do so, we consider a natural generalization of the commonly-used Susceptible-Infected-Recovered (SIR) model on networks in which some individuals wear masks while others do not. We allow for different probabilities of transmission between mask-wearers and non-mask-wearers, so that an individual wearing a mask is less likely to be infected. We refer to this model as the {\it mask model}. For networks with a given degree distribution, we provide analytical methods to accurately predict the total number of infected individuals of each type (mask-wearing and non-mask-wearing) as well as the probability that an epidemic will emerge. Technically, this is achieved by adapting techniques developed by Alexander and Day \cite{alexander2010risk} as well as Eletreby, Zhuang, Carley, Ya\u{g}an and Poor \cite{eletreby2020effects}, which were used to study a multi-strain model with mutation. Finally, we conduct extensive simulations to illustrate how mask-wearing can impact the spread of an epidemic.

\subsection{Related Work}

Classical models of epidemics use a system of ordinary differential equations (ODEs) to describe the fraction of susceptible, infected and recovered individuals within the population (see for instance \cite{brauer2012mathematical}). Prior models which incorporate the effects of mask-wearing have modified the basic ODE model in various ways. Brienen et al \cite{brienen_masks} considered a simple modification in which the reproductive number of the virus, $R_0$, is reduced by a multiplicative factor based on the efficacy of masks. Subsequently, Tracht et al \cite{tracht_masks} as well as Eikenberry et al \cite{eikenberry_masks} considered more complex generalizations of the basic ODE model in which mask-wearers and non-mask-wearers have different transmissibilities and mask-wearers become non-mask-wearers at some rate, as well as vice versa. While ODE-based models are relatively simple to simulate and analyze, they are only mathematically justified under the unrealistic assumption that an infected individual can transmit the virus to {\it any} other susceptible individual in the population, regardless of location or other factors. 

Our approach, on the other hand, falls under the class of {\it network epidemic models}. These models take an individual-level view of viral spread, and studies how the structure of the contact network influences the epidemic. This provides much finer information about the epidemic, but is costly to simulate, spurring a large body of work devoted to deriving {\it analytical predictions} of epidemiological properties \cite{Newman_2002, moore2000exact,meyers2007contact}. In particular, our work is closely related to literature on heterogenous bond percolation \cite{allard2009heterogenous,lee2020epidemic} and a multiple-strain model with mutations \cite{alexander2010risk, eletreby2020effects}. We elaborate on these connections in later sections.

\section{EPIDEMIC MODELS}

The most basic model of network epidemics was studied by Newman \cite{Newman_2002}; we briefly review his setup in order to provide context for the more complex models we consider in this paper. Given a prescribed degree distribution (for instance Poisson or Power law), a random contact network is generated via the configuration model \cite{molloy1995critical,Bollobas,newman2001random}. Initially, a single individual (patient zero) is infected with the virus, and each neighbor of patient zero becomes infected with probability $T$, where $T$ is referred to as the {\it transmissibility} of the virus. Patient zero then recovers and is no longer susceptible. The process continues as each newly-infected vertex attempts to infect their susceptible neighbors in the same manner. The process terminates when there are no more susceptible vertices in the population.

\subsection{Single-strain propagation with masks}

To account for the effects of mask-wearing on viral spread, we make the following modifications to Newman's model. First, we specify $m \in [0,1]$ to be the expected fraction of individuals who wear a mask. Formally, we assign each vertex in the contact network a mask with probability $m$ and no mask with probability $1 - m$. This is done independently for each vertex. Second, we assume that the transmissibilities are {\it heterogenous}: the probability that individual $u$ infects individual $v$ depends on whether $u$ and $v$ are wearing masks. We say that a vertex is of type 1 if they wear a mask and type 2 if they do not wear a mask. We then have four parameters describing the transmissibility of the virus: $T_{11}, T_{12}, T_{21}$ and $T_{22}$. The parameter $T_{11}$ is the transmissibility when $u$ and $v$ both wear masks, $T_{12}$ is the transmissibility when $u$ wears a mask and $v$ does not, etc. For brevity, we refer to this model as the {\it Mask Model}. This type of model is sometimes called heterogenous bond percolation over multi-type networks. We remark that while Allard et al \cite{allard2009heterogenous} consider such a model in full generality, an important contribution of this paper is to study in detail the important case of mask-wearing. After the initial submission of our paper in September 2020, Lee and Zhu \cite{lee2020epidemic} studied the same Mask Model we propose here and derived the epidemic threshold and expected epidemic size using similar techniques as Allard et al. Here, using different techniques we additionally characterize the probability of emergence and provide extensive simulations to support our results.

\subsection{Multi-strain Model with Mutation}

In \cite{alexander2010risk}, Alexander and Day proposed a multiple-strain model that accounts for mutations between strains. In their model, there are $d$ possible strains of a virus with transmissibilities given by $Q_1, \ldots, Q_d$. If an individual is infected with strain $i$, the virus may mutate into a different strain within the host. Formally, the probability that strain $i$ mutates into strain $j$ within a host is given by $\mu_{ij}$. 

We next describe a mapping between the Mask Model and the multi-strain model with mutation. The key insight is that in expectation, a mask-wearing individual will have a different effective transmissibility than a non-mask-wearing individual. This will allow us to map the mask-wearing model into a two-strain model with mutation. 

We begin by deriving the transmissibilities of the two-strain model. Suppose that a vertex $v$ is infected and wears a mask. Since each neighbor wears a mask with probability $m$, the expected transmissibility of $v$ is given by 
\begin{equation}
\label{eq:Q1}
Q_1 : = T_{11} \cdot m + T_{12} \cdot (1 - m).
\end{equation}
Similarly, if $v$ does not wear a mask, the transmissibility is given by 
\begin{equation}
\label{eq:Q2}
Q_2 : = T_{21} \cdot m + T_{22} \cdot (1 - m). 
\end{equation}
Proceeding with the analogy, the mutation probability $\mu_{11}$ is the fraction of mask-wearing neighbors infected by a mask-wearer. This is given by 
\begin{equation}
\label{eq:mu11}
\mu_{11} : = \frac{ T_{11} \cdot m}{T_{11} \cdot m + T_{12} \cdot (1 - m)}.
\end{equation}
Using the same reasoning, we can compute the other three mutation probabilities as 
\begin{align}
\label{eq:mu12}
\mu_{12} & = \frac{ T_{12} \cdot (1 - m)}{ T_{11} \cdot m + T_{12} \cdot (1 - m)} \\
\label{eq:mu21}
\mu_{21} & = \frac{T_{21} \cdot m}{T_{21} \cdot m + T_{22} \cdot (1 - m)} \\
\label{eq:mu22}
\mu_{22} & = \frac{T_{22} \cdot (1 - m)}{T_{21} \cdot m + T_{22} \cdot (1 - m) }.
\end{align}
The advantage of this formulation is that it allows us to compute analytical predictions for the probability of emergence and epidemic size in the Mask Model using the methods of Eletreby et al \cite{eletreby2020effects}, which hold for the multi-strain model with mutation. 

\section{ANALYSIS}

In this section, we derive analytical predictions for the probability of emergence and the expected epidemic size. One way to do so is by formulating the mask model as a multi-strain model with mutation and then leverage the analytical predictions of Eletreby et al \cite{eletreby2020effects}. We also compute the probability of emergence and epidemic size {\it directly} for the mask model, using methods developed by Alexander and Day \cite{alexander2010risk} as well as Eletreby et al \cite{eletreby2020effects}.

\subsection{Probability of emergence}
Emergence refers to the event where the epidemic process persists over time and keeps infecting susceptible individuals. Extinction, on the other hand, is the event where the epidemic dies out in finite time. In this section, we show how to compute $P_1$ (resp. $P_2$) which is the probability of extinction given that patient zero wears a mask (resp. does not wear a mask). The probability of emergence can then be computed as $1 - P_1$ if patient zero wears a mask and $1 - P_2$ otherwise. 

Our analysis follows the method of Alexander and Day \cite{alexander2010risk}, who derived expressions for the probability of emergence in the multi-strain model with mutation. Suppose that a randomly chosen vertex $v$ is patient zero and assume that $v$ wears a mask. Let $X$ (resp. $Y$) be the number of mask-wearing (resp., non-mask-wearing) neighbors of $v$ who are infected by $v$. Then, conditioned on $v$ having $k_1$ susceptible mask-wearing neighbors and $k_2$ susceptible non-mask-wearing neighbors, $X$ and $Y$ are independent with $X \sim \mathrm{Binomial}(k_1, T_{11})$ and $Y \sim \mathrm{Binomial}(k_2, T_{12})$. Thus for $s, t \in \R$, 
\begin{multline*}
\E [ s^X t^Y \mid k_1, k_2] \\
= (1 - T_{11} + s \cdot T_{11})^{k_1} (1 - T_{12} + t \cdot T_{12} )^{k_2}.
\end{multline*}
Next, if $k$ is the total number of susceptible neighbors, we have $k_1 \sim \mathrm{Binomial}(k, m)$ and $k_2 = k - k_1$. Hence
\begin{multline*}
\E [ s^X t^Y \mid k] = \sum\limits_{k_1 = 0}^k {k \choose k_1}  \left(  m ( 1 - T_{11} + s T_{11}) \right)^{k_1} \\
\times \left(  (1 - m) (1 - T_{12} + t T_{12} )\right)^{k - k_1} \\
= \left( 1 - (m T_{11} + (1 - m) T_{12}) + (ms T_{11} + (1-m) t T_{12} ) \right)^k.
\end{multline*}
Using the analogy with the multi-strain model with mutation (see \eqref{eq:Q1}-\eqref{eq:mu22}) we can equivalently write
\begin{equation}
\label{eq:conditional_pdf}
\E [ s^X t^Y \mid k] = (1 - Q_1 + Q_1 (s \mu_{11} + t \mu_{12}))^k.
\end{equation}
We remark that \eqref{eq:conditional_pdf} also holds in the multi-strain model with mutation \cite[Section 2.2]{alexander2010risk}, implying the probability of emergence is {\it identical} in both models. For completeness, we describe how to compute this probability.

Define $\gamma_1(s,t)$ to be the probability generating function (PGF) for the number of infections of each type (mask-wearing or not) emanating from patient zero. We have
$$
\gamma_1(s,t) = g( 1 - Q_1 + Q_1 (s \mu_{11} + t \mu_{12}) ),
$$
where $g$ is the PGF of the degree distribution, i.e., $g(z) = \sum_{k = 0}^\infty p_k z^k$. Following the same arguments, we have 
$$
\gamma_2(s,t) = g( 1 - Q_2 + Q_2 (s \mu_{21} + t \mu_{22}) ).
$$
The PGF of the number of infections of each type emanating from a later-generation infective wearing a mask, given by $\Gamma_1(s,t)$, is
$$
\Gamma_1(s,t) = G( 1 - Q_1 + Q_1 (s \mu_{11} + t \mu_{12})),
$$
where $G$ is the PGF for the {\it excess} degree distribution, i.e., $G(z) = \sum_{k = 0}^\infty \frac{k p_k }{ \langle k \rangle} z^k$. We also have
$$
\Gamma_2(s,t) = G( 1 - Q_2 + Q_2 (s \mu_{21} + t \mu_{22})).
$$
With the derived PGFs in hand, the probability of extinction starting from a later-generation infective, given by $q_1$ (resp., $q_2$) if patient zero wears a mask (resp., does not wear a mask), is the smallest non-negative solution of the fixed-point equation $(s,t) = (\Gamma_1(s,t), \Gamma_2(s,t))$. Finally, the probability of emergence starting from patient zero, denoted by $P_1$ (resp., $P_2$) if patient zero wears a mask (resp., does not wear a mask) is given by $(P_1, P_2) = (\gamma_1(q_1, q_2), \gamma_2(q_1,q_2) )$.

Since the probability of emergence is the same in the Mask model and the multi-strain model with mutation, the critical threshold at which an epidemic emerges (also known as the reproductive number $R_0$) is the same in both models as well. To calculate this threshold, we first introduce two matrices: 
$$
\mathbf{Q} : = \begin{pmatrix} Q_1 & 0 \\ 0 & Q_2 \end{pmatrix} \qquad \text{and} \qquad \boldsymbol{\mu} := \begin{pmatrix} \mu_{11} & \mu_{12} \\ \mu_{21} & \mu_{22} \end{pmatrix}.
$$
Then the formula for the critical threshold \cite{eletreby2020effects,alexander2010risk} is
$$
R_0 : = \left( \frac{\langle k^2 \rangle - \langle k \rangle}{\langle k \rangle} \right) \rho (\mathbf{Q} \boldsymbol{\mu}),
$$
where $\rho(\mathbf{Q} \boldsymbol{\mu} )$ denotes the spectral radius of $\mathbf{Q} \boldsymbol{\mu}$ and $\langle k \rangle, \langle k^2 \rangle$ are the first and second moments of the degree distribution, respectively. If $R_0 < 1$ the epidemic dies out in finite time, and if $R_0 > 1$ an epidemic persists. To write things in terms of the parameters of the Mask model, we can define the matrices 
$$
\mathbf{T} : = \begin{pmatrix} T_{11} & T_{12} \\ T_{21} & T_{22} \end{pmatrix} \qquad \text{and} \qquad \mathbf{m} : = \begin{pmatrix} m & 0 \\ 0 & 1-m \end{pmatrix},
$$
and note that $\mathbf{Q} \boldsymbol{\mu} = \mathbf{Tm}$. Hence we equivalently have
$$
R_0 = \left( \frac{\langle k^2 \rangle - \langle k \rangle }{ \langle k \rangle} \right) \rho(\mathbf{Tm}).
$$

\subsection{Expected epidemic Size}

We follow the method of Eletreby et al \cite{eletreby2020effects}. Since the contact network $G$ is drawn from the configuration model with degree distribution $\{p_k \}_k$, it is locally tree-like. We can compute the probability that a given vertex is infected by considering the tree-like neighborhood around it. Mathematically, we can consider an infinite rooted tree where the bottom level is labeled level zero and the top (the root) is labeled level infinity. We let $q_{\ell,1}$ (respectively, $q_{\ell,2}$) be the probability that a mask-wearing (respectively, non-mask-wearing) vertex is infected in level $\ell$. 

The pair $(q_{\ell + 1,1}, q_{\ell + 1,2} )$ can be recursively computed from $(q_{\ell,1}, q_{\ell,2} )$ as follows. Consider a vertex in level $\ell + 1$ that wears a mask. It has degree $k$ with probability $\frac{k p_k }{\langle k \rangle}$ due to properties of the configuration model. Due to the tree structure, $k-1$ of these edges are sent to $\ell$ and one is sent to the parent in level $\ell + 2$. Out of the $k-1$ level-$\ell$ neighbors, some number $X$ wear a mask while the rest do not, where $X \sim \mathrm{Binomial}(k-1,m)$. Out of the $X$ mask-wearing neighbors, some number $U$ are infected, where $U \sim \mathrm{Binomial}(X,q_{\ell,1})$. Similarly, there are $V$ infected non-mask-wearing neighbors, where $V \sim \mathrm{Binomial}(k - 1 - X, q_{\ell,2})$. Finally, if there are $U$ infected mask-wearing neighbors and $V$ infected non-mask-wearing neighbors, the probability that the mask-wearing parent in level $\ell + 1$ becomes infected is $1 - (1 - T_{11})^U (1 - T_{21})^V$. If we define 
\begin{multline}
\label{eq:f1}
f_1(z,q_1,q_2) : = \sum\limits_{x = 0}^z {z \choose x} m^x (1 - m)^{z - x} \\
\times \sum\limits_{u = 0}^x {x \choose u} q_1^u (1 - q_1)^{x - u} \\
\times \sum\limits_{v = 0}^{z - x} {z - x \choose v} q_2^v (1 - q_2)^{z - x - v} \\
\times (1 - (1 - T_{11})^u (1 - T_{21})^v ),
\end{multline}
then we have 
$$
q_{\ell + 1,1} = \sum\limits_{k = 0}^\infty \frac{k p_k }{\langle k \rangle} f_1 (k-1, q_{\ell, 1}, q_{\ell, 2} ).
$$
If we define $f_2$ to be the same as \eqref{eq:f1} except the term 
$$(1 - (T - T_{11})^u (1 - T_{21})^v )$$
is replaced by 
$$(1 - (1 - T_{12})^u (1 - T_{22})^v ),$$
then we also have
$$
q_{\ell + 1,2} = \sum\limits_{k = 0}^\infty \frac{k p_k }{\langle k \rangle} f_2(k - 1, q_{\ell, 1}, q_{\ell, 2} ).
$$
Following the analysis in \cite{eletreby2020effects}, the sequence $\{q_{\ell,1}, q_{\ell,2}\}_{\ell \ge 1}$ converges to a limit $(q_{\infty,1}, q_{\infty,2})$ which satisfies the fixed point equation
$$
q_{\infty,i} = \sum\limits_{k = 0}^{\infty} \frac{k p_k }{ \langle k \rangle} f_i (k - 1,q_{\infty, 1}, q_{\infty,2} ), \qquad i \in \{1,2\}.
$$
Finally, to compute the probability of infection at the root, we note that the root has $k$ neighbors with probability $p_k$, and all neighbors are in a lower level. Thus, if $S_1$ ($S_2$) is the probability of infection of a mask-wearing (non-mask-wearing) root vertex, then we have
$$
S_i = \sum\limits_{k = 0}^\infty p_k f_i(k, q_{\infty,1}, q_{\infty,2}), \qquad i \in \{1,2\}.
$$
As our analysis was for an arbitrarily chosen root node, $S_i$ is the expected fraction of mask-wearing vertices that eventually get infected by the epidemic, conditioned on the epidemic occurring. The total fraction of infections is then given by $S = S_1 \cdot m + S_2\cdot (1 - m)$. 

\section{NUMERICAL RESULTS}

\subsection{Epidemic as a function of the mean degree}

\begin{figure}[htp]
    \centering
    \includegraphics[width=0.45\textwidth]{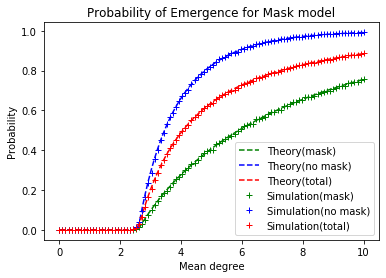}
    \caption{Plot of the probability of emergence from simulations and theoretical predictions. The degree distribution is Poisson with varying mean.}
    \label{fig:pe}
\end{figure}

We conducted extensive numerical simulations to validate our theoretical analysis. In Figure \ref{fig:pe}, we study the probability of emergence. The contact network was generated via the configuration model with Poisson degree distribution and 500,000 vertices. We studied several values for the mean degree ranging between 0 and 10. To generate the simulation plots, we took an average over 20,000 independent trials where, in each trial, a new contact network was generated. The parameters of the mask model were chosen to be $m = 0.45$, $T_{11} = 0.126, T_{12} = 0.18, T_{21} = 0.42, T_{22} = 0.6$. The choice of $m$ was based on the current fraction of mask-wearers in the US \cite{ihme}. The transmissibility parameters were chosen as a reasonable baseline to illustrate the model and our theoretical results about the model. For larger mean degrees, we see that we have a near-perfect match between the simulations and theoretical predictions. For smaller mean degrees, the match is close, but not perfect, since the emergence event becomes quite rare close to the phase transition point. We expect that if much larger networks are used, the simulations will enjoy better alignment with the theoretical predictions, even close to the phase transition point. 

\begin{figure}[htp]
    \centering
    \includegraphics[width=0.45\textwidth]{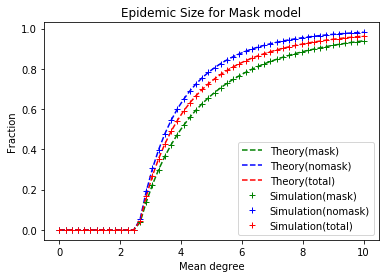}
    \caption{Plot of the expected epidemic size conditioned on emergence from simulations and theoretical predictions, with Poisson degree distribution. The empirical and theoretical curves match very well, even close to the phase transition point.}
    \label{fig:es}
\end{figure}

In Figure \ref{fig:es}, we study the expected size of the epidemic, conditioned on emergence. In our simulations, we used the same number of nodes and degree distribution, averaged over 10,000 independent trials. The same parameters for the mask model were used as well. We see very good alignment between the simulations and theoretical predictions, confirming the validity of our theoretical results.

\begin{figure}[htp]
    \centering
    \includegraphics[width=0.45\textwidth]{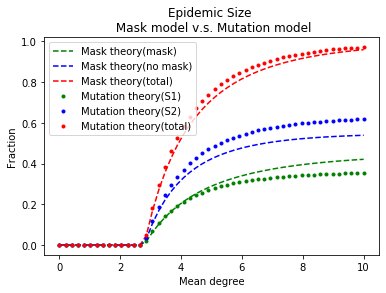}
    \caption{Comparison of the theoretical prediction for the expected epidemic size conditioned on emergence in the mask model (circles) and the multi-strain model with mutation (dashed lines). There is good alignment close to the critical threshold, but the predictions diverge for larger mean degrees. We use the same parameters as in Figure \ref{fig:es}.}
    \label{fig:mask_mutation_theory}
\end{figure}

In Figure \ref{fig:mask_mutation_theory}, we illustrate the interesting finding that while the multi-strain model with mutations can be used to compute the probability of emergence in the Mask model, it yields an {\it incorrect} prediction of the expected epidemic size. There seems to be a good alignment between the two curves close to the critical threshold, but the two predictions diverge for larger mean degrees. We give a possible reason for this mismatch. In the mask model, there is a single strain in the population and a susceptible vertex is infected as long as as there is a successful infection by at least one neighbor. In the multi-strain model with mutation, if there are multiple successful infections to a susceptible vertex, the resulting transmitted strain depends on the number of successful infections of each type. When the mean degree is small, it is unlikely that there will be more than one successful infection, as the number of neighbors of a vertex is small. However as the mean degree increases, the difference becomes more pronounced. We plan to further investigate the fundamental differences between the mask model and multi-strain model with mutations in future work.

\subsection{Epidemic as a function of the fraction of mask-wearers}

\begin{figure}[htp]
    \centering
    \includegraphics[width=0.45\textwidth]{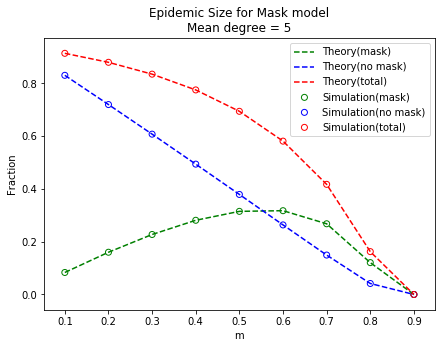}
    \caption{Plot of the expected epidemic size conditioned on emergence, as a function of $m$. While masks decrease the fraction of infections in total (red) and for the non-mask-wearing population (blue), the mask-wearing infections (green) curiously peaks at $m = 0.6$.}
    \label{fig:mask_es}
\end{figure}

Figure \ref{fig:mask_es} illustrates the effect of the probability of mask-wearing, $m$, on the expected epidemic size. In our simulations, we generated the contact network with 5,000,000 vertices and a $\mathrm{Poisson}(5)$ degree distribution. We studied various values of $m$ between 0 and 1. As the fraction of mask-wearing individuals increases, the total number of infections (shown in red) is monotonically decreasing, demonstrating the effectiveness of masks in curbing the spread of COVID-19. Interestingly, we see that the fraction of infected non-mask-wearers (shown in blue) is also monotonically decreasing in $m$. The intuition for this observation is clear; if many individuals wear a mask, on a high level it reduces the effective transmissibility of the virus, thus reducing the number of infected non-mask-wearers as well. Curiously, the fraction of infected mask-wearers is {\it not} monotonically decreasing in $m$; the infection curve peaks at $m = 0.6$. We provide a possible explanation. There are two opposing effects which influence the number of infected mask-wearers. As $m$ increases, the total number of infected mask-wearers will naturally increase, since there are more susceptible mask-wearers in the population. On the other hand, increasing $m$ will also decrease the transmissibility of the virus, leading to a lower rate of infection. When $m < 0.6$, the first effect dominates: the increase in susceptible mask-wearers is greater than the decrease in transmissibility. The point $m = 0.6$ is where the two effect balance each other; for $m > 0.6$, the decrease in transmissibility dominates the increase in susceptible mask-wearers. 

\begin{figure}[htp]
    \centering
    \includegraphics[width=0.45\textwidth]{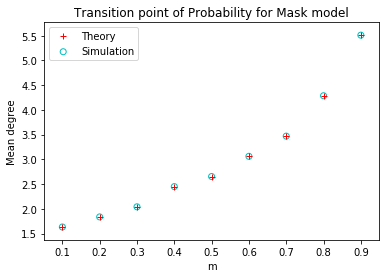}
    \caption{Plot of the simulated and theoretical critical degree for an epidemic to persist. In our simulations, we generated a contact network with Poisson degree distribution on 5,000 vertices. We averaged over 1,000 independent simulations to generate the simulation data points for the plot.}
    \label{fig:mask_pe}
\end{figure}

In Figure \ref{fig:mask_pe}, we study how the critical threshold depends on $m$. As one may expect, as the fraction of mask-wearers increase, a larger mean degree is required for an epidemic to emerge.

\subsection{Epidemic as a function of the baseline transmissibility}

\begin{figure}[htp]
    \centering
    \subfigure[]{
    \includegraphics[width=0.45\textwidth]{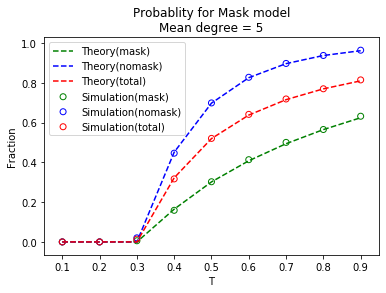}}
    \subfigure[]{
    \includegraphics[width=0.45\textwidth]{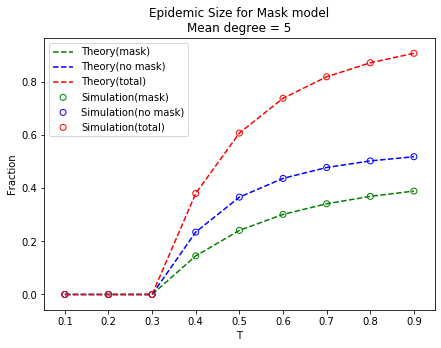}}
    \caption{
Empirical and theoretical plots for the probability of emergence (a) and expected epidemic size (b) as a function of $T$. The green (blue) curves assume that patient zero is wearing (not wearing) a mask. The red curve assumes that patient zero is randomly selected.}
    \label{fig:T_pe_es}
    \vspace{-3mm}
\end{figure}

In Figure \ref{fig:T_pe_es}, we conside the effect of the baseline transmissibility (i.e., the transmissibility between two non-mask-wearers) on the probability of emergence and expected epidemic size. Instead of setting specific values for the transmissibilities in the mask model, we assume that masks have an inward efficiency of $T_{mask,1}$ and an outward efficiency of $T_{mask,2}$. This implies that the transmission parameters have the form $T_{11} = T_{mask,1} T_{mask,2} T$, $T_{12} = T_{mask,2} T$ and $T_{22} = T$. Here, we fix $T_{mask,1} = 0.3$ and $T_{mask,2} = 0.7$ (this reflects the observation that masks have higher outward efficiency than inward) and we study how the epidemic characteristics change with $T$. In our simulations, we set $m=0.45$ and assumed a $\mathrm{Poisson}(5)$ degree distribution and generated networks with 5,000,000 vertices, averaging over 100 independent simulations. In both the probability of emergence and the expected epidemic size, the curves are increasing with $T$, and an epidemic emerges when $T = 0.3$. While the simulated probability of emergence deviates from the theoretical curve, we expect to see concentration as we increase the number of experiments.

\section*{Acknowledgements}
This work was supported in part by 
the National Science Foundation through grants RAPID-2026985, RAPID-2026982, CCF-1813637 and DMS-1811724; the Army Research Office through grants \# W911NF-20-1-0204 and \# W911NF-17-1-0587; and  the C3.ai Digital Transformation Institute. 
\section{CONCLUSION}

In this paper, we studied the effects of mask-wearing on viral spread, specifically the probability of emergence and the expected epidemic size conditioned on emergence. We offered two different perspectives on modeling viral spread with masks: through a heterogeneous bond percolation approach on multi-type networks and through an analogy with a multiple-strain model with mutation. Theoretically, we find that while the probability of emergence is the same in both models, the expected epidemic size can be different. We also show that the expected epidemic size is decreasing as a function of the fraction of mask-wearing individuals, confirming that mask-wearing can be an effective strategy in curbing the spread of COVID-19. 

\addtolength{\textheight}{-12cm}


\bibliographystyle{IEEEtranS}
\bibliography{references}

\end{document}